\documentclass[aps,prd,twocolumn,groupedaddress,nofootinbib]{revtex4}  
\usepackage{graphicx}  
\usepackage{dcolumn}   
\usepackage{bm}        
\usepackage{amssymb,amsfonts,amsmath,physics}   
\usepackage{slashed}
\usepackage{subfig}
\PassOptionsToPackage{demo}{graphicx}
\usepackage{multirow}
\usepackage{framed}
\usepackage{bbm}

\usepackage{tikz-feynman}
\tikzfeynmanset{compat=1.1.0}
\usetikzlibrary{arrows}
\tikzset{ vertex/.style={circle,draw, minimum size=1.5em}, edge/.style={->,> = latex'}}

\usepackage[bookmarks, breaklinks, colorlinks, urlcolor=blue, citecolor=blue, linkcolor=blue]{hyperref}

\usepackage[normalem]{ulem}

\newif\ifstartedinmathmode
\newcommand\encircled[1]{%
  \relax\ifmmode\startedinmathmodetrue\else\startedinmathmodefalse\fi%
  \tikz[baseline,anchor=base]{%
  \node[draw,circle,outer sep=0pt,inner sep=.2ex]
    {\ifstartedinmathmode$#1$\else#1\fi};}%
}

\def\to {\rightarrow}

\newcommand{\bmt}{\begin{pmatrix}}
\newcommand{\emt}{\end{pmatrix}}
\newcommand{\ba}{\begin{array}{c}}
\newcommand{\ea}{\end{array}}
\newcommand{\beq}{\begin{equation}}
\newcommand{\eeq}{\end{equation}}
\newcommand{\bea}{\begin{eqnarray}}
\newcommand{\eea}{\end{eqnarray}}

\newcommand{\bi}{\begin{itemize}}
\newcommand{\ei}{\end{itemize}}

\newcommand{\baz}{\begin{array}{cc}}
\newcommand{\mathsym}[1]{{}}

\newcommand{\bt}{\begin{tabular}}
\newcommand{\et}{\end{tabular}}

\newcommand{\benu}{\begin{enumerate}}
\newcommand{\eenu}{\end{enumerate}}
\newcommand{\bpm}{\begin{pmatrix}}   
\newcommand{\epm}{\end{pmatrix}}

\def\q2 {q^2}

\def\lum{{\mathfrak{L}}_{\tt int} }

\def\zBB{{\mathbbm Z}}
\def\ccal{\mathcal{C}}
\def\ocal{\mathcal{O}}
\def\su#1{{SU(#1)}}
\def\ui{U(1)}

\def\bin{{\sf s}}
\def\ctw{c_{2\tt w}}

\def\hide#1{}

\begin{document}


\title{Optimal New Physics estimation in presence of Standard Model backgrounds}

\author{Subhaditya Bhattacharya$^{a}$}
\email[E-mail: ]{subhab@iitg.ac.in}

\author{Sahabub Jahedi$^{a}$}
\email[E-mail: ]{sahabub@iitg.ac.in}

\author{Jayita Lahiri$^{b}$} 
\email[E-mail: ]{jayita.lahiri@desy.de}

\author{Jose Wudka$^{c}$} 
\email[E-mail: ]{jose.wudka@ucr.edu}
\vspace{2cm}

\affiliation{
	{\color{blue}$^{a}$}Department of Physics, Indian Institute of Technology Guwahati, Assam, 781039, India}

\affiliation{{\color{blue}$^{b}$}Institut f{\"u}r Theoretische Physik, Universit{\"a}t Hamburg, 22761 Hamburg, Germany}

\affiliation{{\color{blue}$^{c}$}Department of Physics and Astrophysics, University of California, Riverside, California 92521, USA}


\begin{abstract} 
In this work, we develop a numerical technique for the optimal estimation of the new physics (NP) couplings applicable to any collider process 
without any simplifying assumptions. This approach also provides a way to measure the quality of the NP estimates 
derived using standard $ \chi^2$ analysis, and can be used to gauge the advantages of various modalities of collider design. We illustrate the 
techniques and arguments by considering the pair production of heavy charged fermions at an $e^+e^-$ collider.
\end{abstract}

\maketitle

\noindent
The search for New Physics (NP) relies to a great extent on estimating or constraining its couplings in collider observables. 
The best fit values of a coupling is usually obtained using a likelihood ($\chi^2$) function that can also be used to obtain a 
confidence interval for that coupling. The optimal observable technique (OOT) \cite{Diehl:1993br,Gunion:1996vv,Bhattacharya:2021ltd} 
is well suited to study the quality of such confidence regions, since it can be used to determine the minimum statistical uncertainty regions 
for the couplings of interest \footnote{Our discussion will deal solely with statistical uncertainties as the OOT does not provide an analysis tool for systematic errors.}
\cite{Grzadkowski:1996pc,Grzadkowski:1997cj,Gunion:1998hm,Grzadkowski:1998bh,Grzadkowski:1999kx,Grzadkowski:2000nx,Hagiwara:2000tk,Grzadkowski:2004iw,Grzadkowski:2005ye,Hioki:2007jc,Dutta:2008bh,Jahedi:2022duc,Bhattacharya:2023mjr,Jahedi:2023myu,Bhattacharya:2023beo,Jahedi:2024kvi,Jahedi:2024wnw}. In collider experiments (to which we restrict ourselves in this paper), a realistic comparison of the confidence regions obtained using the 
likelihood function and the OOT approach must include all relevant SM background contributions, which can significantly affect the outcome. 

The main goal of this paper is to provide a practical method that allows the application of the OOT to any process in any collider environment, without any  approximations. We then provide a tool for using these results to determine the collider parameters (such as beam polarization, luminosity etc.) that can enhance the sensitivity to a specific type of NP. The results obtained using the method here developed are also compared to those obtained using standard likelihood techniques, which provides a quantitative measure of the quality of this last, more traditional, approach. As far as the authors are aware, no similar analysis has appeared previously in the literature. 

%
%

When including background contributions, the observable cross sections of interest take the general form,
\beq
\frac{d\sigma}{d\phi} = \ocal(\phi) =  \frac{d\sigma_{\tt sig}}{d\phi} + \frac{d\sigma_{\tt bkg}}{d\phi} = \sum_i g_i^0 f_i(\phi) \,,
\label{eq:obs}
\eeq
where $\phi$ denotes all phase-space variables. The signal contribution $d\sigma_{\tt sig}$ includes all NP effects, 
while the contribution to $d\sigma_{\tt bkg}$ are assumed known.
We will be interested in measuring (or bounding) the parameters $ g_i^0 $, which are known functions of the coupling 
constants (NP and SM); the functions $ f_i(\phi) $ are linearly-independent and also known~\footnote{Choice of $ g_i^0$ 
is not unique, but the $\chi^2_{\tt t}$ function derived below is.}.

In the OOT \cite{Gunion:1996vv}, one defines a normalized probability density $ p(\phi)=\ocal/\sigma $, 
where $\sigma$ is the total cross section, and observables $\gamma_i $ by
\beq
\gamma_i =  \left(M^{-1}\right)_{ij} \frac{f_j}p \,;\quad M_{ij} = \int \frac{f_i(\phi) f_j(\phi)}{\ocal(\phi)} d\phi\;;
\eeq
where the integration region is restricted by all experimental cuts relevant to the reaction under consideration. 
The average of $\gamma_i$ (using $p$) equals to $g_i^0 $, and their correlation matrix equals $ V = M^{-1}/\lum $, 
where $\lum$ denotes the integrated luminosity of the collider experiment. These observables are optimal in the sense 
that their statistical uncertainty is minimal, so that for another set of observables $ \gamma'_i$ whose average is also 
$g_i^0$ and have correlation matrix $V'$, $ {\bf a}^T V^{-1}{\bf a} \le {\bf a}^T V'{}^{-1}{\bf a} $ for all vectors {\bf a}. 
Using $V$ we define,
\beq
\chi^2_{\tt t}=\sum_{i,j}(g_i-g^0_i)(g_j-g^0_j)V_{ij}^{-1};\quad V = \frac{1}{\lum}   M^{-1}\,;
\label{eq:chi2.oot}
\eeq
that can be used to determine the confidence regions in parameter space given the central, or ``seed'' values $ g_i^0$. 

The OOT has been generally used for the cases where the SM background can be ignored; an analytical expression for the 
hard process is obtained and an efficiency factor introduced to mimic experimental cuts and final-state effects such as branching ratios \cite{Gunion:1996vv,Bhattacharya:2021ltd}. 
Unfortunately this approximation is not always reliable, nor can it be systematically improved. In order to sidestep these problems we 
develop below a straightforward numerical procedure for calculating $V$ including all final-state effects and experimental cuts without 
approximation, and which can be applied to any reaction irrespective of the strength of the SM background, and in any collider environment.
This feature allows the application of this approach to any collider search, for any type of NP (supersymmetry, effective theories, 
neutrino physics, dark matter, etc.)

We first sketch the methodology for a simple hypothetical scenario, which can be easily generalized to more complicated situations. 
Suppose the theory under study has two NP parameters of interest, $a, b$, and the amplitude is linear in these parameters, 
so that the differential cross-section of the final state event takes the form,
\beq
\ocal=f_0 + af_1 + b f_2 +a^2 f_3 + ab f_4 +b^2 f_5 =\sum^{5}_{i=0}g^0_i f_i,
\label{eq-obs-sp}
\eeq 
where we follow the notation used in Eq.~\eqref{eq:obs}; the $i=0$ term with $g_0^0=1$ and $f_0$ represents the pure SM contribution, including 
non-interfering SM background contributions. Now, we divide the phase space integration region into $\tt R$ bins $\bin_1,~\bin_2,...,\bin_{\tt R}$ and define, 
\beq
\Delta_{\tt r}=\int_{\bin_{\tt r}}d\phi, ~~~~n_{{\tt r}i}=\lum \int_{\bin_{\tt r}}d\phi~ f_i\,.
\eeq
Then total number of events in $\bin_{\tt r}$ is
\beq
N_{\tt r}=\lum \int_{\bin_{\tt r}} d\phi ~\ocal=\sum_{i} g^0_i n_{{\tt r}i}.
\eeq
If the bins are small enough, $f_i$ will be approximately constant inside them, whence 
\begin{align}
\begin{split}
f_i|_{\bin_{\tt r}} \simeq \frac{1}{\Delta_{\tt r}} \int_{\bin_{\tt r}} d\phi f_i= \frac{1}{\Delta_{\tt r}}\frac{n_{{\tt r}i}}{\lum},\\
\ocal|_{\bin_{\tt r}} \simeq \frac{1}{\Delta_{\tt r}} \int_{\bin_{\tt r}} d\phi~ \ocal= \frac{1}{\Delta_{\tt r}}\frac{N_{\tt r}}{\lum}\,.\\
\end{split}
\end{align}
Therefore,
\beq
M_{ij}=\sum_{r}\int_{\bin_{\tt r}} d\phi \frac{f_i f_j}{\ocal} \simeq \sum_{r} \frac{1}{\lum \Delta_{\tt r}} \frac{n_{{\tt r}i}n_{{\tt r}j}}{N_{\tt r}}\,.
\eeq

Finally, noting that $N_{\tt r}$ depends on $a,\,b$ only through the coefficients $ g_i^0$ whose expressions are known, 
one can extract the $n_{{\tt r}i}$ from $N_{\tt r}(a,b)$ for a few values\footnote{For the present case of two NP parameters, six such values are needed; for $q$ NP parameters $1+q(q+3)/2 $ values are required.} of $a,\,b$; explicitly
\begin{align}
\begin{split}
n_{{\tt r}0}=&N_{\tt r}(0,0) \,,\\
n_{{\tt r}1}=&\frac{1}{2}[-3N_{\tt r}(0,0)+4N_{\tt r}(1,0)-N_{\tt r}(2,0)] \,,\\
n_{{\tt r}2}=&\frac{1}{2}[-3N_{\tt r}(0,0)+4N_{\tt r}(0,1)-N_{\tt r}(0,2)]\,,\\
n_{{\tt r}3}=&\frac{1}{2}[N_{\tt r}(0,0)-2N_{\tt r}(1,0)+N_{\tt r}(2,0)]\,,\\
n_{{\tt r}4}=&N_{\tt r}(0,0)-2N_{\tt r}(0,1)-N_{\tt r}(1,0)+N_{\tt r}(1,1)\,,\\
n_{{\tt r}5}=&\frac{1}{2}[N_{\tt r}(0,0)-2N_{\tt r}(0,1)+N_{\tt r}(0,2)]\,.\\
\label{eq:nri}
\end{split}
\end{align}

$N_{\tt r}(a,b)$ can be obtained with all the experimental cuts implemented 
in event simulations, leading to a straightforward evaluation of 
$M$ and thus optimal $\chi^2_{\tt t}$ as in Eq.~\ref{eq:chi2.oot}, whose accuracy increases with the number of bins used. 
 Note that the efficiency in background reduction (having smaller $f_0$ in Eq.~\eqref{eq-obs-sp}) via judicious selection cuts and beam polarization
helps more precise optimal estimation of NP couplings ($a,b$).

\begin{figure}[htb!]
	$$
	\includegraphics[height=3.7cm,width=6.5cm]{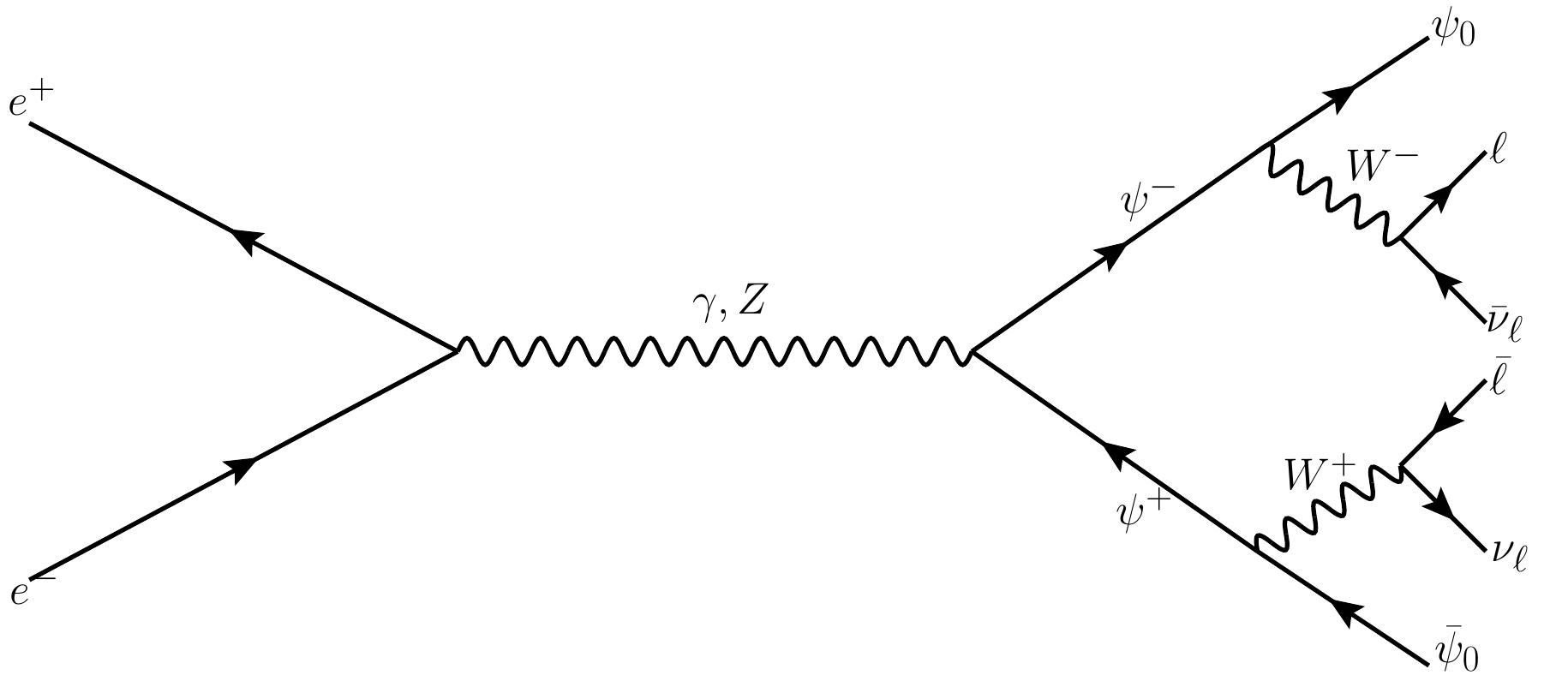}
	$$
	\caption{\small Pair-production of heavy charged fermions ($\psi^+\psi^-$) and their subsequent decay at an $e^+ \, e^-$ collider.}
	\label{fig:productn}
\end{figure}

We illustrate the method using a simple example of a dark matter (DM) model.
The NP consists of heavy charged and neutral fermions, 
$ \psi^\pm$ and $ \psi_0$, respectively, having $ \psi_0 \psi^\pm W^\mp$ coupling. These particles appear in extensions 
of the SM with a fermion isodoublet $\Psi = \left(\psi_1,\psi^-\right)^T$ having hypercharge $Y_\Psi=-1$ and a singlet $\chi$ 
\cite{Bhattacharya:2015qpa}, providing a viable DM candidate $\psi_0$ (a mass eigenstate), when they are 
stabilized by a discrete $\mathbbm{Z}_2$ symmetry. Here we take on a purely phenomenological approach, considering $ \psi^\pm $ to have general 
chiral couplings to the $Z$ boson parameterized by two NP couplings, $a$ and $b$, but minimal coupling to photon \cite{Bhattacharya:2021ltd}, 
\beq
\psi^+\psi^-Z:- ~\frac{i e_0}{2 s_w c_w}\gamma^\mu\left(a+b\gamma^5\right)\;,~\psi^+\psi^-\gamma:- ~ie_0\gamma^\mu\,,
\label{eq:vertex1}
\eeq
where $e_0$ is $U(1)_{\tt em}$ coupling constant, $s_w$ and $c_w$ are the cosine and sine of the weak mixing angle, respectively. See Appendix \ref{sec:model} for 
details of the model and related DM phenomenology.

We now calculate the optimal ({\it i.e.} minimal) statistical uncertainty when measuring the NP couplings $a,\,b$ in the pair production of $ \psi^\pm$ (with their subsequent decays) at an $e^+e^-$ collider (see Fig.~\ref{fig:productn}): 
\begin{equation}
e^+e^- \rightarrow \psi^+\psi^- \rightarrow  W^+  W^- \psi_0 \bar\psi_0 ~, ~W\to \ell\nu  \,,
\label{eq:reaction}
\end{equation}
with $ \ell = e,\,\mu$. We assume the $\psi_0(\bar{\psi_0})$ to be stable DM candidates that escapes detection, so that the signal consists 
of missing energy ($\slashed E=\sqrt{s}-E_{\rm vis}$) plus opposite sign dileptons (OSL)~\footnote{A similar analysis has been done in \cite{Bhattacharya:2021ltd}, 
but without including the SM background.}. We include all possible 2-body and 3-body SM backgrounds: 
$ e^+e^- \to WW,\,ZZ,\,\mu\mu,\,\tau\tau,\,WWZ,\,\ell\ell Z $, with the subsequent leptonic decays of the $W,\,Z$ and $ \tau $. 
In order to suppress the SM backgrounds we impose thee following cuts~\footnote{The ideal choice of cuts 
(those that minimize the area of the $ \chi^2_{\tt t}\le1$ regions) are model-dependent; we will not study here the optimization of such cuts, 
restricting ourselves to reasonable choices based on physical considerations.}:
\begin{itemize}
	\item $\ccal_1$: $p_T^{\ell} > 10$ GeV, $N_{\tt lep}=2$, $ |m_{\ell\ell} - m_Z| > 15$ GeV and $ \Delta R_{\ell\ell} < 3.0$,
	\item $\ccal_2$: $\slashed{E} > 325$ GeV\,,
\end{itemize}
\begin{figure*}[htb!]
	\includegraphics[scale=0.3]{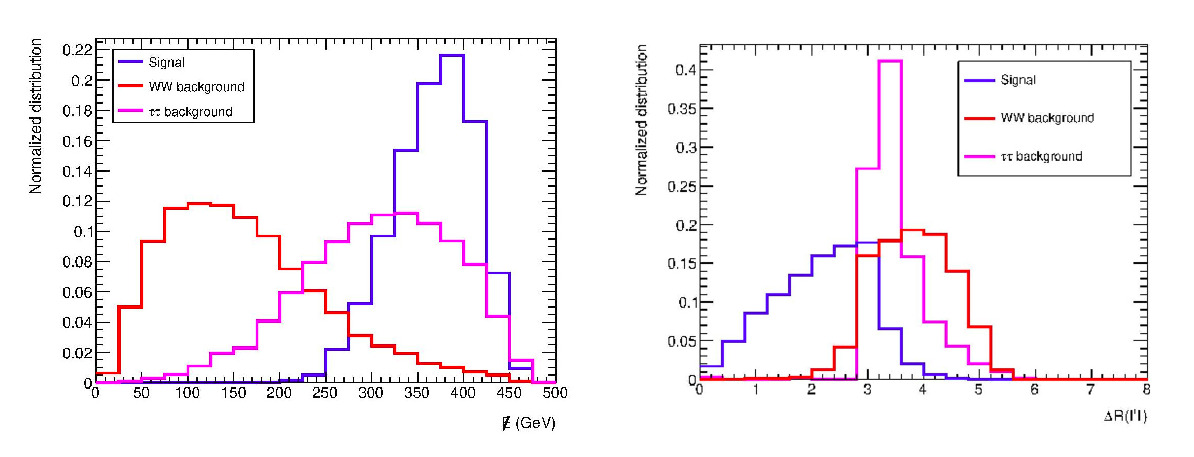}
	\caption{\small Normalized $\slashed{E}$ distribution (left) and $\Delta R(\ell^+\ell^-)$ (right) for signal and dominating backgrounds of the 
		reaction in Fig. \ref{fig:productn}; see text for details. We assumed $\sqrt{s} = 500$ GeV, $m_{\psi^{\pm}} = 210$ GeV 
		(consistent with LEP bound \cite{L3:2001xsz}) and $m_{\psi_1} = 60$ GeV.}
	\label{fig:sigbkg}
\end{figure*}
where $N_{\tt lep}$ is the number of light charged leptons in the final state, $m_{\ell\ell}$ the invariant mass of the two final charged leptons, 
and $\Delta R = \sqrt{(\Delta \eta)^2+((\Delta \phi)^2)}$ is the angular separation between two opposite sign leptons. 
$\ccal_2$ reduces the (dominant) $W$ background, while $\ccal_1$ suppresses the $Z,\,\mu$ and $ \tau $ backgrounds; see Fig.~\ref{fig:sigbkg}. 

We now determine the accuracy to which $a$ and $b$ can be measured, assuming these couplings have one of the following seed values: 
$\bullet$ $a^0=0$, $b^0=1$ (pure axial coupling), $\bullet$ $a^0=1$, $b^0=0$ (pure vector coupling), $\bullet$ $a^0=1$, $b^0= 1$ (chiral coupling).
%
\begin{figure*}[hbt!]
	\includegraphics[height=4cm,width=15cm]{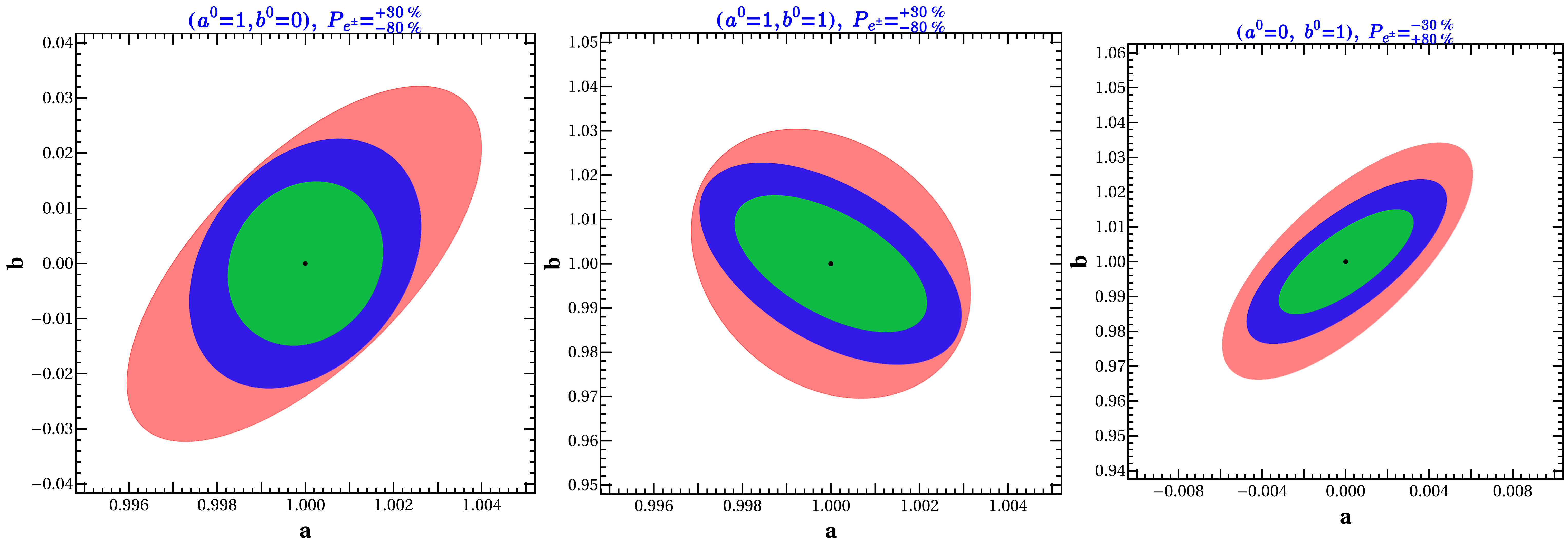}
	\caption{\small Effects of cuts on the 1$\sigma$ surfaces for 3 different hypotheses: signal only with $\ccal_1+\ccal_2$ applied (green); and signal+background after  $\ccal_1$ (light read), and  $\ccal_1+\ccal_2$  (blue) -- see the text for details.}
	\label{fig:1sigmapoleps}
\end{figure*}
We list in table~\ref{bkgs} the results for both signal and 
background event numbers, and the corresponding production cross sections (note that for this reaction there is no SM-DM interference); 
the results illustrate the effectiveness of  the cuts imposed in reducing the background, and also the effects of the beam polarization. 
We assumed the the collider CM energy is $\sqrt{s}=500$ GeV, integrated luminosity ${\mathfrak{L}_{\tt int}} = 2000~\rm{fb^{-1}}$, and $ m_{\psi^\pm}=210\,\text{GeV},\, m_{\psi_0}=60\,\text{GeV}$.

The choice of beam polarization plays a crucial role in optimizing signal to noise ratio, that in turn helps in reducing the uncertainty of the coupling estimates. 
For example, the  choice $P_{e^\pm} = ^{+30\%}_{-80\%}$ (consistent with the ILC design \cite{Behnke:2013xla}), increases the signal cross-section 
when $a^0=1,\,b^0=0$  (it also increases the $WW$ background, but this is suppressed by $\ccal_2$); this results in smaller uncertainties than those 
obtained using unpolarized beams. For $a^0=0,b^0=1$ the choice $P_{e^\pm} = ^{-30\%}_{+80\%}$  enhances the signal cross-section, and also 
suppresses the $WW$ background. Details of signal and background cross-sections, in particular, the dependence on beam polarisation is provided in the Appendix~\ref{sec:SMbackg}.

\begin{table}[!hptb]
	\begin{center}
				\setlength{\tabcolsep}{0.9pt}
		\begin{tabular}{|c|c|c|c|c|}
			\hline
			\multirow{ 2}{*}{Processes} & \multicolumn{2}{c|}{Production cross-section} & \multicolumn{2}{c|}{Number of events}\\
			 & \multicolumn{2}{c|}{after $\ccal_1$ (fb)} & \multicolumn{2}{c|}{after $\ccal_1$ and $\ccal_2$} \\
			\hline  
			\multicolumn{1}{|c}{} &  \multicolumn{1}{|c|}{$P_{e^\pm} = ^{-30\%}_{+80\%}$}  & \multicolumn{1}{c}{$P_{e^\pm} = ^{+30\%}_{-80\%}$} & \multicolumn{1}{|c}{$P_{e^\pm} = ^{-30\%}_{+80\%}$} & \multicolumn{1}{|c|}{$P_{e^\pm} = ^{+30\%}_{-80\%}$} \\ 
			\hline  
			\multicolumn{1}{|c}{$a^0=1,b^0=0$} &  \multicolumn{1}{|c|}{5.2}  & \multicolumn{1}{c|}{55.7} & \multicolumn{1}{c|}{4791} & \multicolumn{1}{c|}{50832} \\ 
			\hline
			\multicolumn{1}{|c}{$a^0=0,b^0=1$} &  \multicolumn{1}{|c|}{19.6}  & \multicolumn{1}{c|}{20.4} & \multicolumn{1}{c|}{17685} & \multicolumn{1}{c|}{18479} \\ 
			\hline
			\multicolumn{1}{|c}{$a^0=1,b^0=1$} &  \multicolumn{1}{|c|}{7.0}  & \multicolumn{1}{c|}{58.2} & \multicolumn{1}{c|}{6457} & \multicolumn{1}{c|}{53250} \\ 
			\hline
			\multicolumn{1}{|c}{$WW$} &  \multicolumn{1}{|c|}{51}  & \multicolumn{1}{c|}{798} & \multicolumn{1}{c|}{1558} & \multicolumn{1}{c|}{18030} \\ 
			\hline  
			\multicolumn{1}{|c}{$\tau\tau/\mu\mu$} &  \multicolumn{1}{|c|}{57}  & \multicolumn{1}{c|}{68} & \multicolumn{1}{c|}{286} & \multicolumn{1}{c|}{360} \\ 
			\hline  
			\multicolumn{1}{|c}{$ZZ$} &  \multicolumn{1}{|c|}{8.8}  & \multicolumn{1}{c|}{18.9} & \multicolumn{1}{c|}{21} & \multicolumn{1}{c|}{44} \\ 
			\hline 
			\multicolumn{1}{|c}{$\nu\nu Z$} &  \multicolumn{1}{|c|}{3.4}  & \multicolumn{1}{c|}{50} & \multicolumn{1}{c|}{72} & \multicolumn{1}{c|}{1190} \\ 
			\hline 
			\multicolumn{1}{|c}{$\ell\ell Z$} &  \multicolumn{1}{|c|}{16.5}  & \multicolumn{1}{c|}{22.4} & \multicolumn{1}{c|}{18} & \multicolumn{1}{c|}{4} \\ 
			\hline 
			\multicolumn{1}{|c}{$WWZ$} &  \multicolumn{1}{|c|}{0.063}  & \multicolumn{1}{c|}{0.87} & \multicolumn{1}{c|}{21} & \multicolumn{1}{c|}{248} \\ 
			\hline 
		\end{tabular}
		\caption{\small Production cross sections with $ \ccal_1$ imposed and final event numbers after cuts $ \ccal_1+ \ccal_2$, 
		for signal and background contributions to the reaction in Fig. \ref{fig:productn}. We took CM energy $\sqrt{s}=500$ GeV,  
		integrated luminosity ${\mathfrak{L}_{\tt int}} = 2000~\rm{fb^{-1}}$ and two polarization choices; we assumed 
		$m_{\psi^{\pm}} = 210$ GeV and $m_{\psi_0} = 60$ GeV.}
		\label{bkgs}
	\end{center}
\end{table}

\begin{figure*}[htb!]
	\includegraphics[height=4cm,width=15cm]{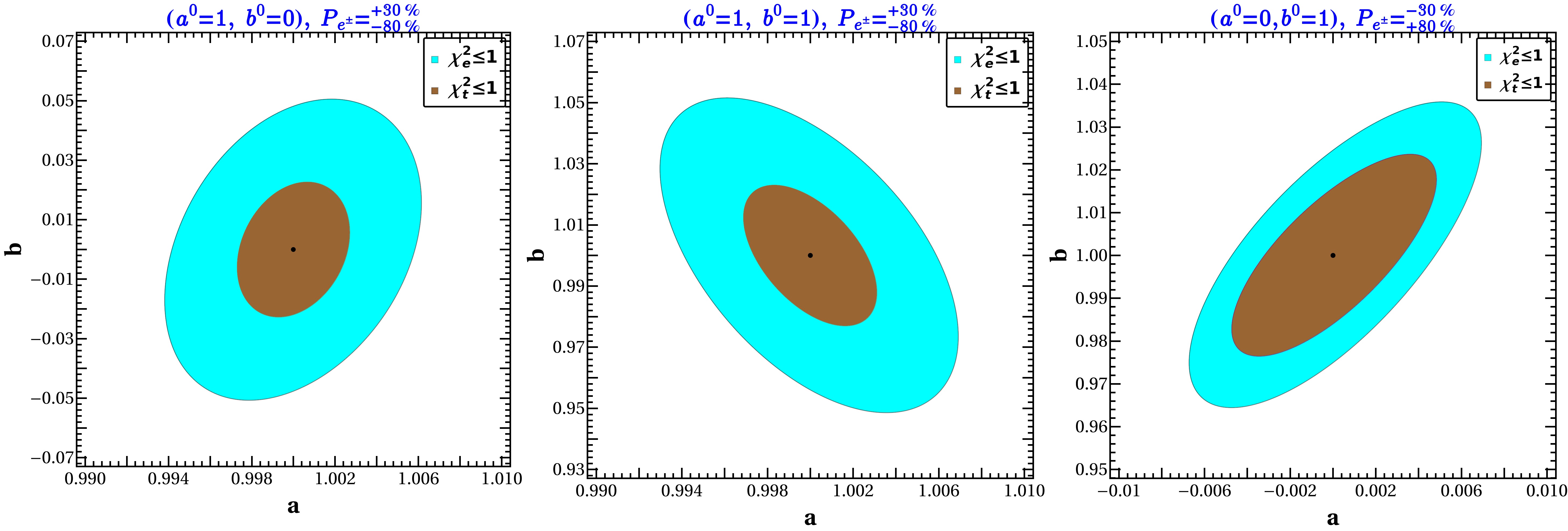}
	\caption{\small Comparison of 1$\sigma$ regions between standard approach (cyan) and OOT (brown) for three different hypotheses.}
	\label{fig:ootvscol1}
\end{figure*}

The 1$\sigma$ regions, defined by $\chi^2_{\tt t} \le 1 $ ({\it cf.} Eq. \ref{eq:chi2.oot}), for the three choices of $ a^0,\,b^0$ considered above, are shown in the Fig.~\ref{fig:1sigmapoleps}, for the signal-only with $\ccal_1$ and $\ccal_2$ imposed (green), for the full signal and background after applying $\ccal_1$ only 
(light red), and for the full signal plus background after applying $\ccal_1$ and $\ccal_2$ (blue). This illustrates the degradation of the (optimal) precision to 
which the NP parameters can be estimated once the background is included (a feature absent in several previous analyses that used the OOT). 

It should be noted that the shape and orientation of the 1$\sigma$ ellipses has a complicated dependence on the NP parameters $a^0,\,b^0$, 
as well as on the polarization and the specific choices made for $ \ccal_{1,2}$. These features cannot be mimicked using an efficiency factor multiplying 
the signal and background cross sections. Such a simplification is appropriate when the background is ignorable, or 
when the cuts reduce signal and background events without changing their phase-space correlation. 
It is therefore more reliable to follow the above procedure than the efficiency approximation.

We now turn to the basic question we wish to address, namely, how close does a standard analysis of collider data come to the 
OOT results when extracting parameter uncertainties? This standard approach is, in general, based on the $\chi^2_{\tt e}$ function, defined as,
\begin{equation}
\chi^2_{\tt e} = \sum_k \left(\frac{N^{\text{exp}}_k - N^{\text{theo}}_k(a,b)}{\sqrt{N^{\text{exp}}_k}}\right)^2\,,
\label{eq:chi2.exp}
\end{equation}
\noindent
where $N_k^{\text{theo}}$ denotes the number of events in the $k^{\tt th}$ bin predicted  by the model (after applying the  cuts $\ccal_1$ and $\ccal_2$), and $N^{\text{exp}}$  the corresponding number of events generated by a simulation of the model with parameters $ a^0,\,b^0$.

In Fig.~\ref{fig:ootvscol1}, we present a comparison of the the 1$\sigma$ regions from both OOT and regular $\chi^2_{\tt e}$ analyses for 
three different hypotheses listed above. The figures also provide a measure of the level of improvement needed in the regular $\chi_{\tt e}^2$ 
analysis to reach the minimal uncertainties derived  using the OOT. For $a^0=1,\,b^0=0,1$, the OOT offers over 100\% improvement in estimating the sensitivity of NP couplings, while providing over 50\% improvement for $a^0=0,\,b^0=1$ (this is because in this case the background is the smallest, and the event distribution closely mimics the pure signal). This situation will repeat in all cases where the signal dominates over the background.

\begin{figure}[htb!]
	\centering
	\includegraphics[scale=0.23]{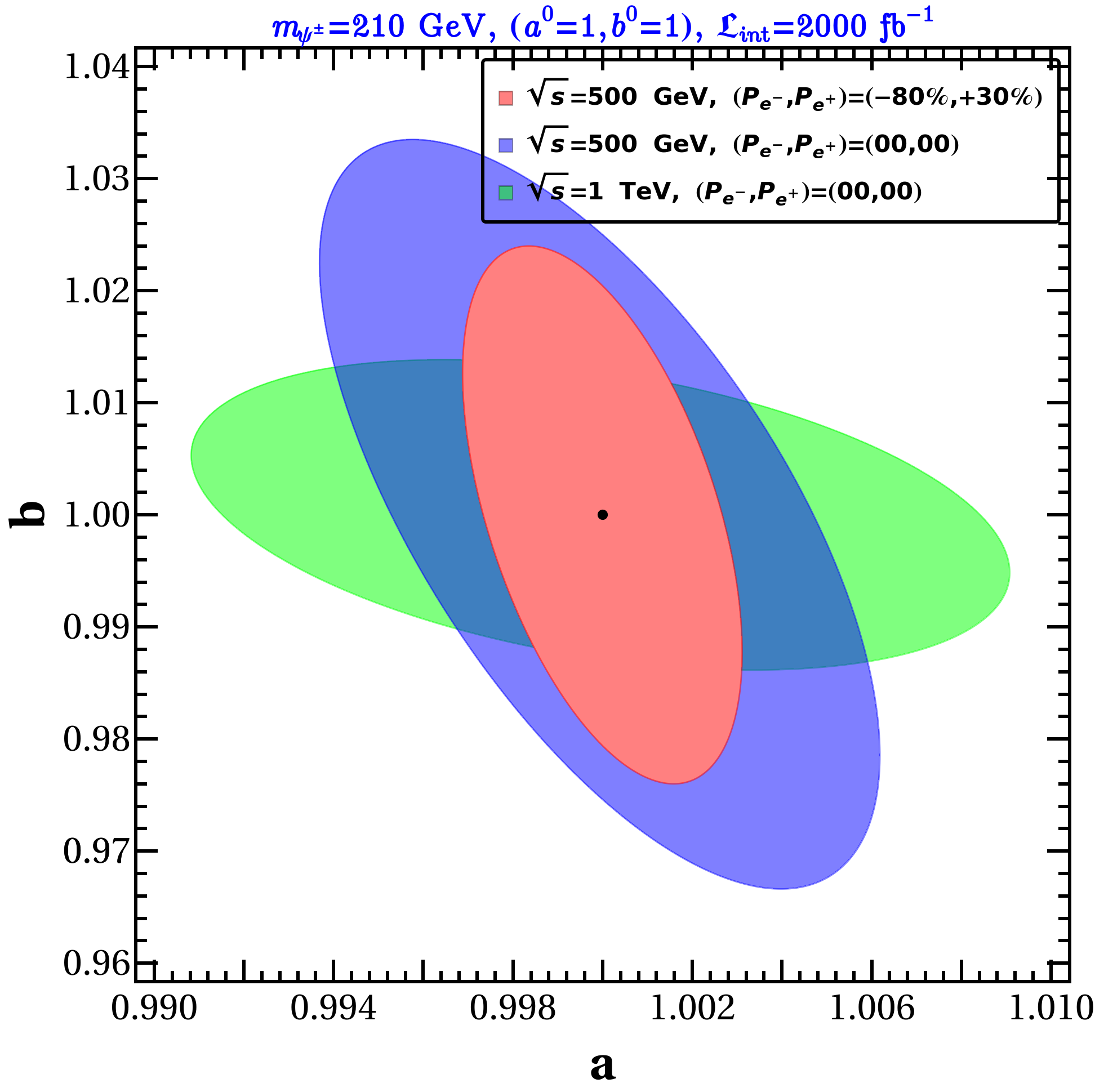}
	\caption{\small Comparison of 1$\sigma$ surfaces for different CM energy and beam polarization.}
	\label{fig:comp}
\end{figure}

It is worthwhile studying how the uncertainty and correlation of NP couplings depend on the CM energy and beam polarization. An illustrative example 
is presented in Fig.~\ref{fig:comp}, from which we can infer that the covariance matrix can be strongly dependent on $ \sqrt{s} $ (for this particular case, 
the uncertainty in $b$ drops, while that in $a$ enhances with larger $\sqrt{s}$); while, as noted previously, the overall uncertainty can be significantly reduced 
by an appropriate choice of polarization. The ability of the OOT to determine the changes in precision to which NP parameters can be extracted with 
both polarization and CM energy provides a useful guide in selecting future collider designs.


\bigskip

In summary, we provide a simple and general numerical method for the application of the OOT to collider processes. This approach allows determination of the optimal statistical precision that can be reached when constraining NP parameters for any model in any given collider process, incorporating all experimental cuts or machine learning techniques to reduce the background. This methodology is also useful in obtaining high-precision estimates of SM parameters ({\it e.g.}~Higgs couplings).

In addition, the results obtained  provide a  gauge of the degree of optimization reached by standard data analysis techniques, and of the effects of collider parameters, such as CM energy and beam polarization (when available) in the precision that can be reached. This last feature can be used to compare different collider proposals, for example, a $\mu^+\mu^-$ collider with relatively high CM energy, 
but no polarized beams, against an $e^+e^-$ collider of lower energy but with beam polarization. Finally, we emphasize that the approach here presented is much preferable to the ``efficiency approximation'' \cite{Gunion:1996vv} where selection cuts and branching ratios are mimicked by multiplying the hard cross sections by a fixed factor. \\

{\bf Acknowledgement} SB acknowledges the grant CRG/2019/004078 from DST-SERB, Govt. of India.

\appendix

\section{Singlet doublet model and dark matter phenomenology}
\label{sec:model}
\noindent
The singlet-doublet dark matter model is one of the simplest extensions of the SM consisting of two additional vector-like leptons: a $SU(2)$ doublet, $\Psi=(\psi_1 ,\, \psi^-\, )$ of hypercharge $Y=-1$, and a singlet $\chi$ of zero hypercharge. Both $ \psi$ and $ \chi $ are assumed odd under a discrete $\zBB_2$ symmetry, while all 
the SM fields are even. Following electroweak symmetry breaking (EWSB), the $\psi\chi H$ Yukawa coupling ({\it cf.} Eq.~(\ref{lag:lagVF}) below) 
induces a mixing of the neutral component $\psi^0$ and $\chi$; the lighter mass eigenstate will serve as a viable DM candidate. The quantum numbers under 
SM$\times \zBB_2$ symmetry are summarized in Table \ref{tab:tab5}.

\begin{table}[htb]
	\begin{center}
		\begin{tabular}{|c||c|c|c|c|}
			\hline
			field & $SU(3)_{\tt C}$ & $SU(2)_{\tt L}$ & $U(1)_{\tt Y}$ & $\zBB_2$ \cr \hline
			$\Psi$ & $1$ & $2$ & $-1$ & $-$ \cr
			$\chi$ & $1$ & $1$ & $0 $ & $-$ \cr \hline
		\end{tabular}
		\caption{\small Quantum numbers of dark-sector fermions under $ \su3_{\tt C} \times \su2_{\tt L} \times \ui_{\tt Y}  \times \zBB_2 $. 
			SM fields are even under $ \zBB_2$.}
		\label{tab:tab5} 
	\end{center}
\end{table}

The Lagrangian of the model is
\begin{align}
\begin{split}
\mathcal{L}^{\tt VF} = &\bar\Psi \left[ i \left(\slashed\partial - i \frac g2 \sigma ^a W_{\mu}^a - i g' \frac{Y}2  B_{\mu}\right)-m_{\psi^\pm} \right]\Psi \\
 &+ \bar\chi \left(i \slashed\partial-m_{\chi} \right)\chi - \left(y\bar\Psi \widetilde{H}\chi + \text{H.c} \right)\,,
\end{split}
 \label{lag:lagVF}
\end{align}
where $H$ is SM Higgs doublet, $B_\mu$ and $W_{\mu}$ are the  $\ui_Y$ and $\su2_L$ gauge fields, and $g,\,g'$ are the corresponding gauge couplings. 
After EWSB, $H$ acquires a $\vev/\sqrt{2}$, and the mass term for dark fermions can be written as,
\begin{align}
-\mathcal{L}_{\tt mass} =  \left( \bar\chi , \bar\psi_1 \right) \bpm m_\chi & \mu \cr \mu & m_{\psi^\pm} \epm \bpm \chi \cr \psi_1 \epm +m_{\psi^\pm}{\psi^+}\psi^-,
\end{align}
where $\mu = y v/\sqrt{2}$. The mass eigenstates $ \psi_{0} $ and $\psi_0^{'}$ are then given by
\beq
\begin{pmatrix}
	\chi \\
	\psi_1
\end{pmatrix} = \begin{pmatrix}
	\cos \gamma & \sin \gamma \\
	-\sin \gamma & \cos \gamma 
\end{pmatrix}  \begin{pmatrix}
	\psi_0 \\
	\psi_0^{'}
\end{pmatrix}; \quad \tan2\gamma = \frac{2\mu}{m_\chi - m_{\psi^\pm}}.
\label{ref:mixang}
\eeq
We will assume~\footnote{The case $ |\mu| < m_{\chi} \ll m_{\psi^\pm} $ is excluded by DM direct-detection and relic abundance constraints.} $ |\mu| \ll m_{\chi} < m_{\psi^\pm} $ so that $2\mu \ll |m_{\chi} - m_{\psi^\pm}|$; in this case $ \gamma $ is small and
\beq
m_{\psi_0}  \simeq  m_\chi - \frac{\mu^2}{m_{\psi^\pm} - m_\chi}\,, \qquad m_{\psi_0^{'}}  \simeq  m_{\psi^\pm} + \frac{\mu^2}{m_{\psi^\pm} - m_\chi};
\eeq
therefore $ m_{\psi_0^{'}} > m_{\psi^\pm}> m_{\psi_0} $ and $ \psi_0$ is the DM candidate. 

\begin{figure*}[htb!]
	\centering
	\includegraphics[scale=0.35]{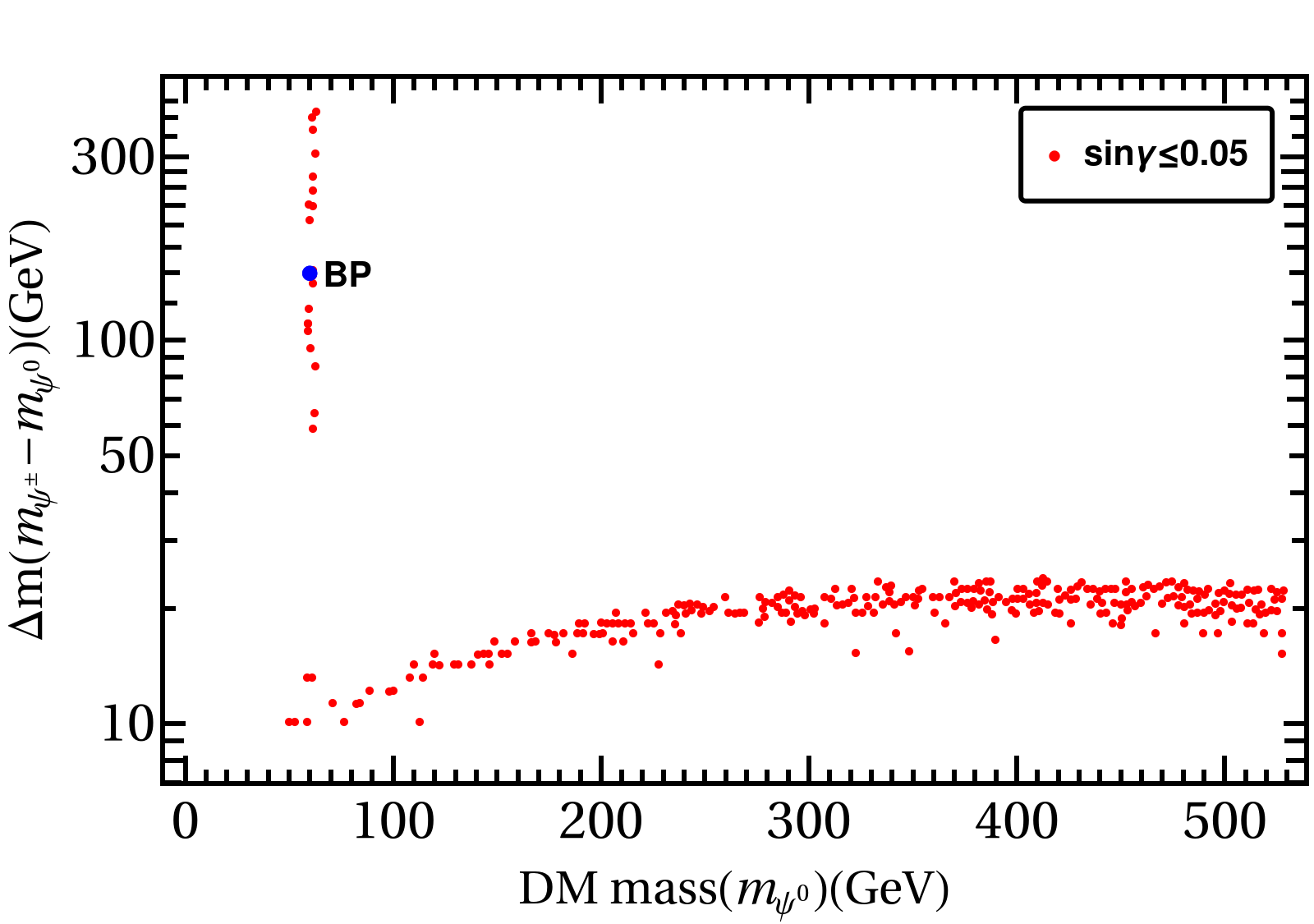}
	\caption{\small Relic density and direct search allowed parameter space in $\Delta m$ vs $m_{\psi_0}$ plane for singlet-doublet dark matter model. The choice 
		of the benchmark point used in the main analysis is indicated.}
	\label{fig:dm.param}
\end{figure*}

We see that the charged heavy fermions (${\psi^\pm}$) have vector-like interactions with $Z$ boson (corresponding $a^0=\ctw\sim1/2,\,b^0=0$) which is similar to one of our hypotheses ($a^0 \ne 0$, $b^0=0$) as described in the main text. $\Delta m$, $m_{\psi_0}$ and $\sin \gamma$ are the free parameters of the model and govern its DM phenomenology. The parameter space in the $\Delta m$ vs $m_{\psi_0}$ plane allowed by relic density and direct search constraints is shown in Fig.~\eqref{fig:dm.param}, along with the choice of a benchmark point that corresponds to the values of $m_{\psi^0,\,\psi^\pm}$ used in the main text. The freedom in choosing $\Delta m$ around $m_{\psi_0}\sim m_h/2\sim 62.5$ GeV, is due to the effects of the Higgs resonance, while in other regions, co-annihilation dependence keeps a tight correlation between $\Delta m$ and $m_{\psi_0}$.

\section{Signal and background cross-sections for different choices of beam polarization}
\label{sec:SMbackg}
\noindent

\begin{figure*}[htb!]
	\centering
	$$
	\includegraphics[scale=0.27]{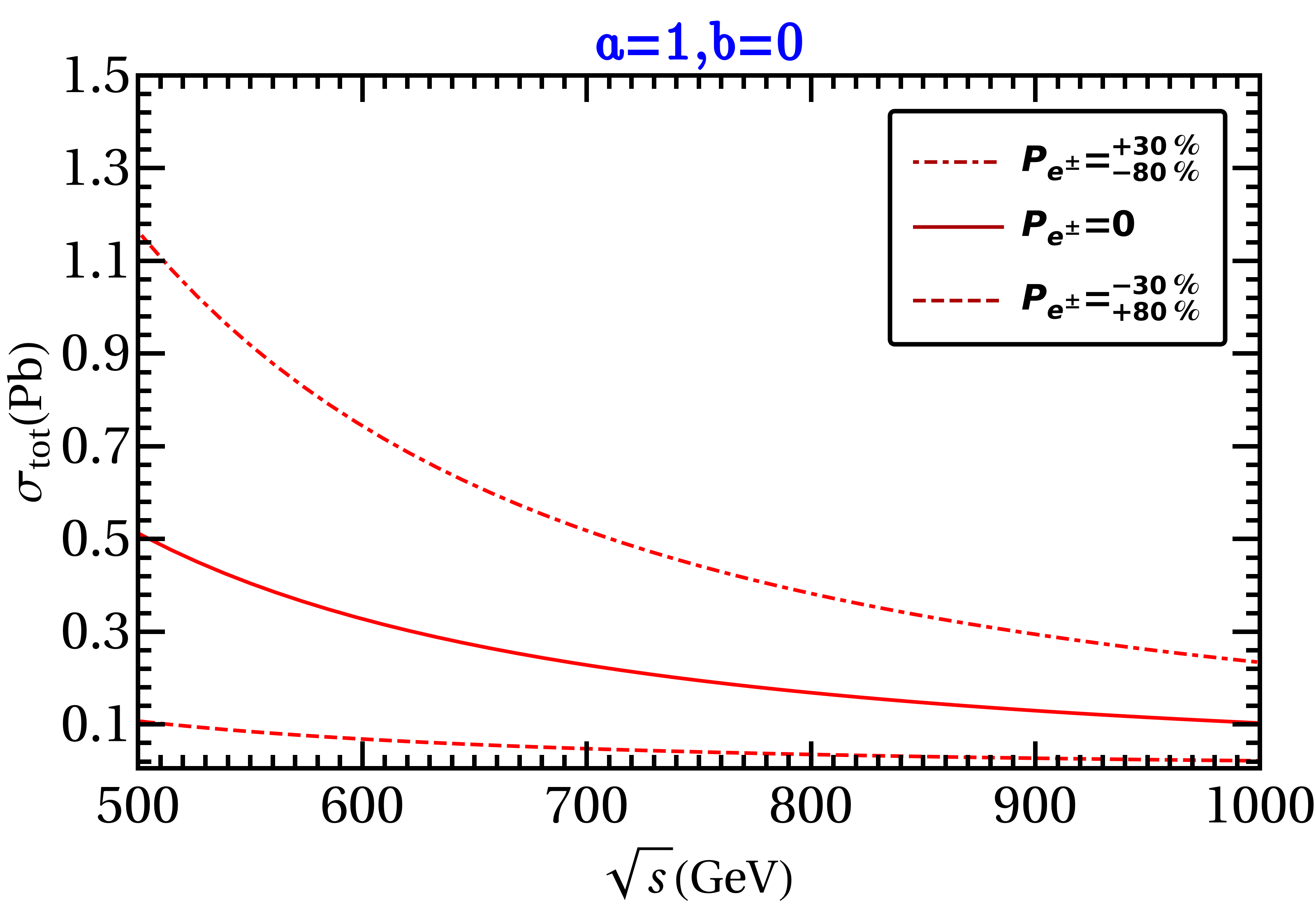}
	\includegraphics[scale=0.27]{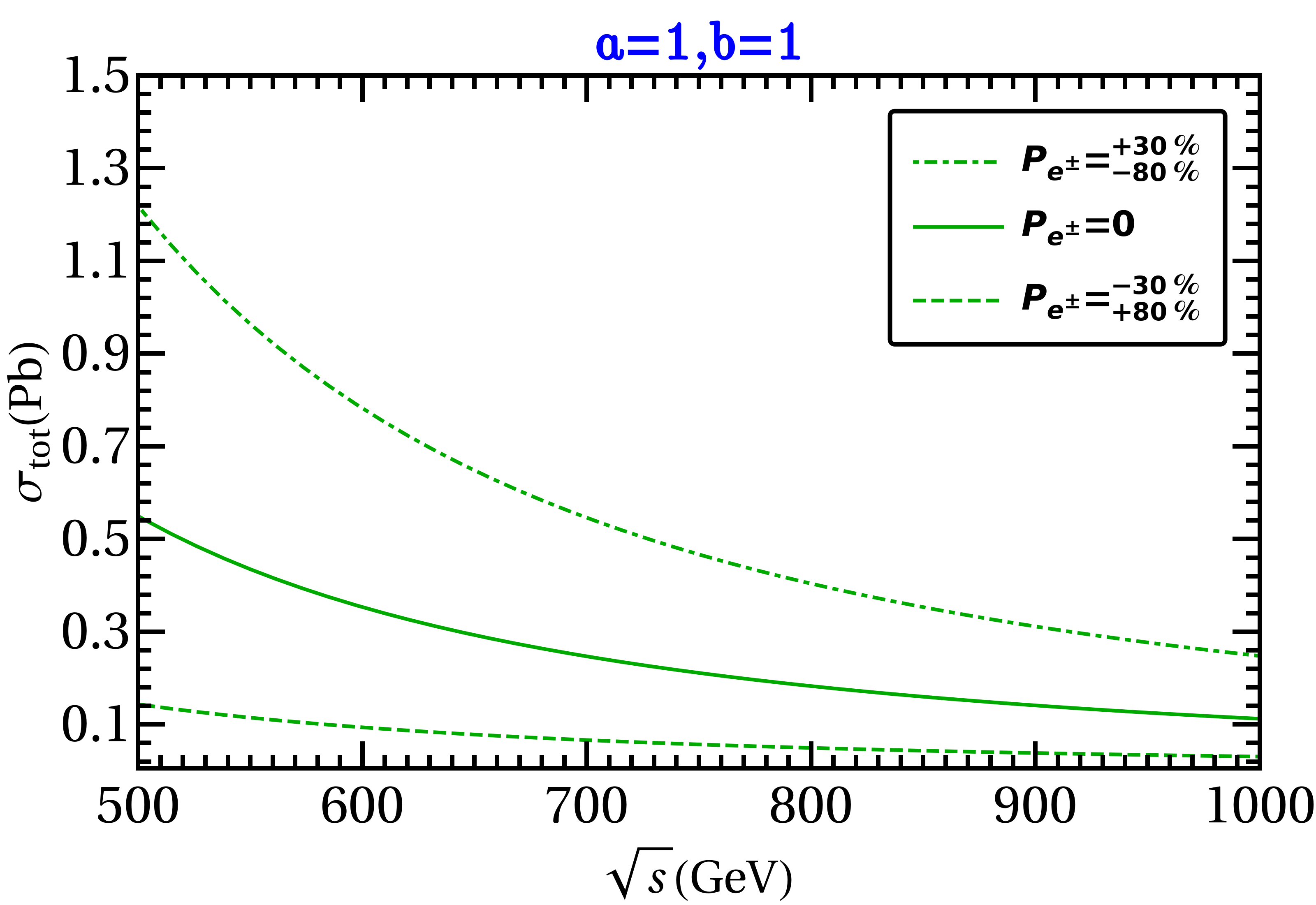}
	\includegraphics[scale=0.27]{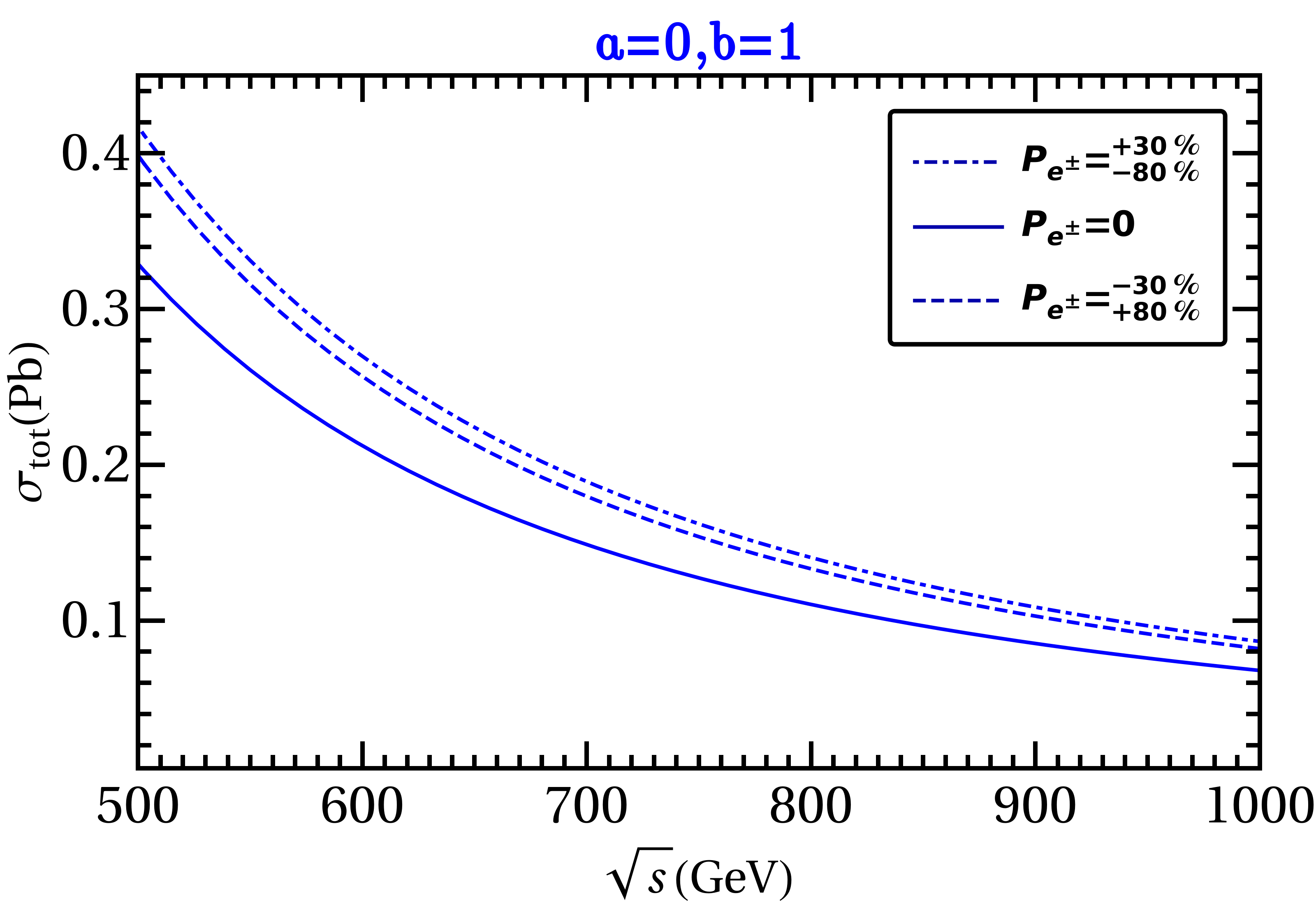}
	$$
	$$
	\includegraphics[scale=0.30]{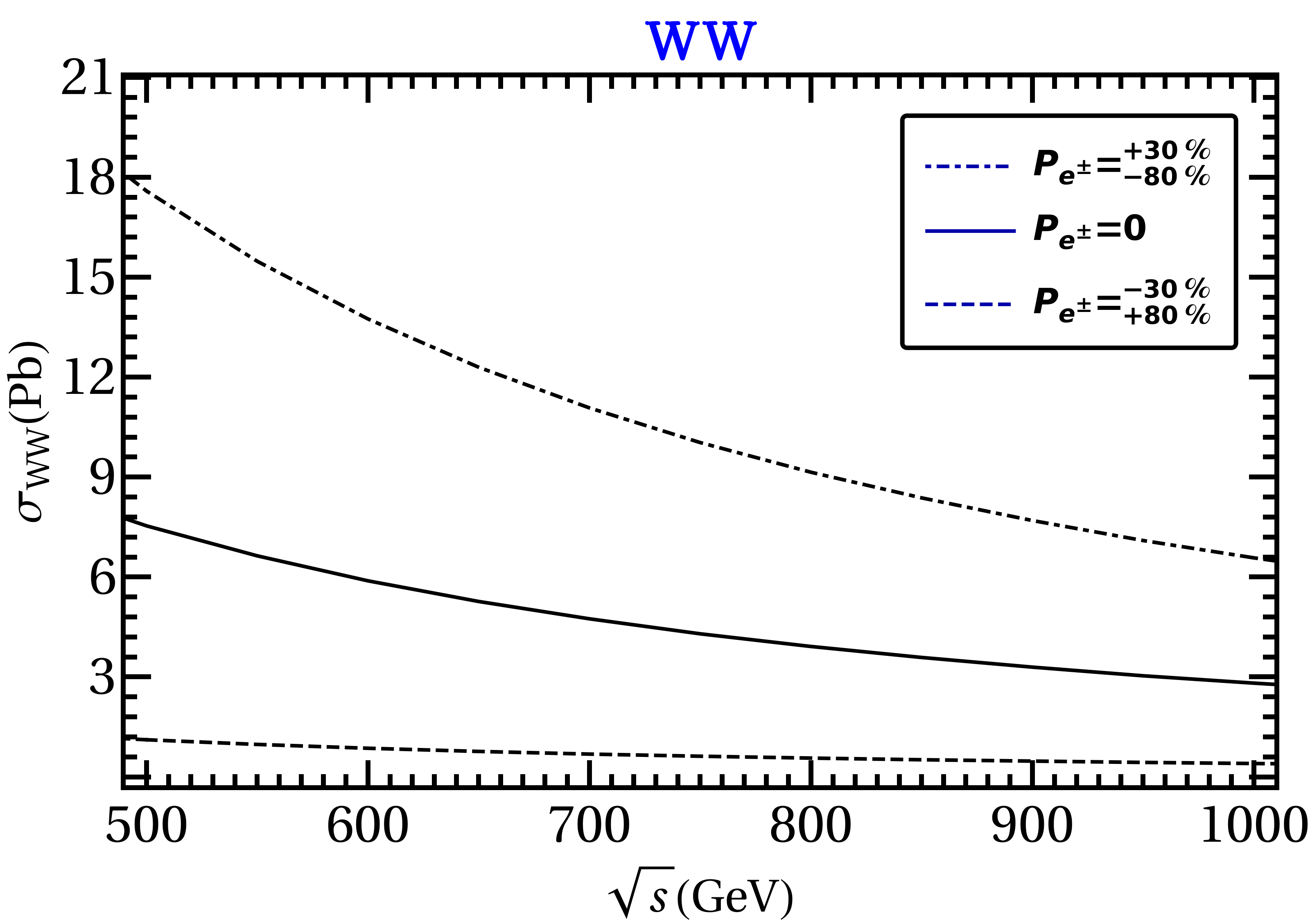}
	$$
	\caption{\small Top: The variation of the total cross-section for the signal with CM energy for different choices of beam polarization. 
		Bottom: Same for the $WW$ background. Beam polarization choices are as indicated in the inset of each figure.}
	\label{fig:xsecab}
\end{figure*}

We now consider the reaction used in the main text: $e^+ e^- \to \psi^+ \psi^- $. The dependence the cross section on $\sqrt{s}$ for three different hypotheses and for different choices of beam polarization is shown in the top panel of Fig.~\eqref{fig:xsecab}. For $a=1,b=0$, $P_{e^\pm} = ^{+30\%}_{-80\%}$ beam polarization combination, enhances the total production cross-section compared to unpolarized beams due to the effects of the vector coupling of $\psi^\pm$ with $Z$. For $a=1,b=1$, the enhancement in the total cross-section is comparatively modest due to the decrease in the total cross-section resulting from the axial vector current. For the purely axial-vector like case  $a=0,b=1$, due to the photon mediation, total cross-section increases while axial-vector current leads to a decrease in cross-section for $P_{e^\pm} = ^{+30\%}_{-80\%}$ polarization combination. If we flip the sign of the beam polarization, the case is reversed. This exemplifies the significance of beam polarization when studying these different models.

The major SM background contribution to the final state signal of two opposite-sign leptons+ missing energy arises from $WW$ production; in $e^+e^-$ colliders this occurs through $s$-channel photon and $Z$, and $t$-channel neutrino exchange. For $P_{e^\pm} = ^{+30\%}_{-80\%}$ combination, the $WW$ cross-section increases compared to unpolarized beam. If we flip the polarization sign, $WW$ cross-section drops, as depicted in the bottom panel of Fig.~\eqref{fig:xsecab}.

\twocolumngrid


\end{document}